
\magnification=1200

\hoffset=0.2truecm 
\voffset=0.5truecm 
\hsize= 5.25in 
\vsize=7.5in 

\def\nl{\par\noindent}

\pretolerance=10000

\font\tenrm=cmr10

\nopagenumbers

\def\({\c c}
\def\|{\'\i}

\baselineskip=14pt 

\rightline { IC/94/305}
\bigskip
\centerline {International Atomic Energy Agency}
\centerline {and}
\centerline {United Nations Educational Scientific and Cultural Organization}
\medskip
\centerline {INTERNATIONAL CENTRE FOR THEORETICAL PHYSICS}
\vskip 1.5truecm
\centerline {\bf LIGHT-FRONT DYNAMICS OF  CHERN-SIMONS SYSTEMS }

\bigskip
\bigskip
\centerline {Prem P. Srivastava\footnote{*}{{\bf Postal address:}
{\it U.E.R.J., Instituto de F\|sica, Rua S\~ao
Francisco Xavier 524,
Rio de Janeiro, RJ, 20550-013, Brasil. }
\nl \quad {\bf E-mail:}\quad prem@cbpfsu1.cat.cbpf.br}}
\bigskip
\nl
\centerline {\it International Centre for Theoretical Physics, Trieste, Italy}
\centerline {and}
\centerline {\it Instituto de F\|sica, Universidade do Estado de Rio de
Janeiro, Rio de Janeiro, Brasil.}
\bigskip
\bigskip
\vskip 0.5cm

\centerline {\bf ABSTRACT}
\bigskip

Chern-Simons theory coupled to complex scalars is quantized on the light-
front in the local light-cone gauge by constructing the self-consistent
hamiltonian theory. It is shown that
no inconsistency arises
on using two local gauge-fixing conditions in the Dirac procedure.
The light-front
Hamiltonian turns out to be simple and the framework may be useful to
construct renormalized field theory of particles
with fractional statistics ({\it anyons}).
The theory is shown to be relativistic and the extra term in the
transformation of the matter field under space  rotations, interpreted in
previous works as anomaly, is argued to be gauge artefact.

\bigskip
\centerline{MIRAMARE- TRIESTE}
\medskip
\centerline { October 1994}
\vfill
\nl PACS number(s): 11.10.Ef, 11.30.Qc, 11.15.Tk

\eject

\baselineskip=18pt  

\headline={\rightheadline}
\def\rightheadline{\tenrm\hfil\folio}
\pageno=1

\nl {\bf 1. Introduction}

\bigskip

Chern-Simons (CS) gauge field theories [1,2] in
three dimensional space time
coupled to matter field have drawn much interest recently. They have been
proposed to describe
excitations with fractional statistics,
{\it anyons}, which have been suggested to be relevant for an explanation [3]
of the fractionally quantized Hall effect and possibly that of
high-$T_{c}$ superconductivity [4] where the dynamics
is effectively confined to a plane. There are, however,
controversies related to the quantized field theoretical formulation.
 The lagrangian (path integral) formulation [5],
 for example, seems to give result which disagree with  the
canonical hamiltonian formulation [6-9].
It is claimed that the theory though shown relativistic has
angular momentum anomaly [10]
or shows anyonicity only in some  nonlocal gauges [9,6].
Internal algebraic inconsistency [9]
of using two  local   gauge-fixing conditions [11]
in the context of the usual Coulomb gauge has also been stressed.
The anomaly is also not found  in some recent works
[12,13] which avoid gauge-fixing
and doubts have been raised about the anyonicity being gauge artefact
[8].

We  scrutinize these points in the paper by quantizing
the CS theory coupled to the complex scalar field
on the light-front [14] in the light-cone gauge. We show that there is
no inconsistency in using two local gauge-fixing conditions in the theory.
It becomes clear that a parallel  discussion is valid
also  in the conventional Coulomb gauge and
even when the fermionic fields are present.
The hamiltonian theory on the light-front
turns out to be simple and practical one and may serve
to construct a renormalized theory in contrast to
the complicated and difficult to handle
hamiltonian obtained in the  equal-time
formulation in the local [6] or nonlocal [9] Coulomb gauge.
The motivation for  the new approach arises from
the recent works  [15,16]
in the light-front quantization showing its potential for
computing  non-perturbative effects in QCD or the study of
relativistic bound state problem. The light-front vacuum is known to be
simpler than
the conventional one and the anyonic
excitations may here be  studied transparently.
In the context of the spontaneous symmetry breaking it was
shown recently [17]
that the light-front formulation leads to the same physical outcome as
that from the equal-time one, though achieved
through a different description.
 The conventional one requires to
{\it add} external constraints in the theory based on physical considerations
while
the similar constraints arise as self-consistency conditions in
the light-front theory. The (anyonic) excitations obeying fractional
statistics would  also emerge in the simpler dynamics [14] on the light-front.
We demonstrate also that the anomaly mentioned above
should rather be interpreted as
gauge artefact here and in the previous works as well.
In the  recently proposed gauge-independent theory [12], where extra
terms are added to the canonical Hamiltonian in ad hoc fashion, it is
not clear if it still describes the original lagrangian theory
and the hamiltonian is as complicated as found in the earlier
works.

In Sec. 2 theory with CS term coupled to  complex scalar field is
quantized on the light-front. The Dirac bracket is
constructed and it is shown to lead to the well known light-front commutators
for the independent scalar fields. The {\it self-consistency } [18] of the
formulation is shown by
recovering the lagrange eqs.
The canonical
Poincar\'e generators are constructed in Sec. 3 and the relativistic
invariance of the theory checked.  The commutators
of the scalar field with Lorentz generators are found
and it is argued  that the so called rotational anomaly in the present
context should rather be  regarded as gauge artefact; with no bearing
on the anyonicity.
\bigskip
\medskip
\nl  {\bf 2. Light-front Quantization of Chern-Simons Theory }

\bigskip

The  CS gauge theory we study
is described by the following lagrangian density

$${{\cal L}} =({{\cal D}}^{\mu}\phi)({\widetilde{\cal D}}_{\mu}\phi^{*})\,+
{\kappa\over {4\pi}}{\epsilon}^{\mu\nu\rho}
A_{\mu} \partial_{\nu} A_{\rho}\eqno(1) $$

\nl Here $\phi$ is a complex scalar filed, $A_{\mu}$ is the gauge field,
${{\cal D}}_{\mu}= (\partial_{\mu}+ie A_{\mu})$,
${\widetilde {\cal D}}_{\mu}= (\partial_{\mu}-ie A_{\mu})$,
$\epsilon^{\mu\nu\rho}
$ is the  Levi-Civita tensor needed to construct the
Chern-Simons kinetic term.
For the coordinates  $ x^{\mu}$, and
for all other vector or tensor quantities, we define the {\it light-front
$\pm$ components} by
$x^{\pm}=(x^{0}{\pm} x^{2})/{\sqrt 2}=x_{\mp}$. We take $x^{+}\equiv \tau $
to indicate the {\it light-front time coordinate} and $ x^{-}$ the
{\it  longitudinal space coordinate} while $x^{1}$ is the {\it transverse }
one.
The metric tensor for the
indices $\mu=(+,-,1)$ is given by the nonvanishing elements
$\,g^{+-}=g^{-+}=-g^{11}=1\,$ and $\epsilon^{+ - 1}= 1\,$.
The  conjugate momenta are

$$\pi={\partial{{\cal L}}\over{\partial(\partial_{+}\phi)}}
={\widetilde {\cal D}}_{-}
\phi^{*}, \qquad \qquad
\pi^{*}={\partial{{\cal L}}\over{\partial(\partial_{+}\phi^{*})}}
={{\cal D}}_{-}
\phi \eqno(2a)$$

$$\pi^{\mu}={\partial{{\cal L}}\over{\partial(\partial_{+}A_{\mu})}}=a\,
\epsilon^{+\mu\nu}A_{\nu}\eqno(2b)$$

\nl where we set ${\kappa= {4 \pi}a}$. They show that we are dealing with a
constrained dynamical system (like
in the equal-time case) and we will
follow the well established Dirac procedure [18]
to construct the hamiltonian theory. We observe that
the  conserved current $j^{\mu}=ie(\phi^{*}{{\cal D}}^{\mu}\phi-
\phi{\widetilde{\cal D}}^{\mu}\phi^{*})\, $  is gauge invariant  and  its
contravariant vector property  should remain  intact
if the theory constructed is relativistic.

  We build the hamiltonian framework in the
	light-cone gauge, $A_{-}\approx 0$.
We recall that it is  mandatory for {\it self-consistency} [18] that the
lagrange eqs.
for the independent fields be recovered from the hamilton's eqs.
We  therefore  examine first the lagrange eqs. in our gauge

$$\eqalignno{ \partial_{+}\partial_{-}\phi &={1\over2}{{\cal D}}_{1}{{\cal
D}}_{1}
\phi -ie A_{+}\partial_{-}\phi-{i\over 2}e (\partial_{-}A_{+})\phi &(3a)\cr
{2a\,}\partial_{-}A_{1}&={j^{+}=ie(\phi^{*}\partial_{-}\phi-
\phi\partial_{-}\phi^{*})} &(3b)\cr
{2a\,}(\partial_{1}A_{+}-\partial_{+}A_{1}) &={j^{-}=
ie(\phi^{*}{{\cal D}}_{+}\phi-
\phi{\widetilde {\cal D}}_{+}\phi^{*})} &(3c)\cr
-{2a\,}\partial_{-}A_{+}&={j^{1}=-ie(\phi^{*}{{\cal D}}_{1}\phi-
\phi{\widetilde{\cal D}}_{1}\phi^{*})} &(3d)\cr}$$

\nl The gauge invariant field $F_{-1}$ reduces in our gauge  to
$\partial_{-}A_{1}$ and it is proportional to the charge density
$j^{+}$. The electric charge is given by $Q=\int d^2x \,j^{+}=2a \int
dx^{1}\,[A_{1}(x^{-}=\infty,x^{1})-A_{1}(x^{-}=-\infty,x^{1})]\,$.
If the  scalar field carries nonzero electric charge,
it follows that  $A_{1}$ may  not
be taken to satisfy the periodic or the vanishing boundary conditions
 along  $x^{-}$ at infinity.
We assume for convenience the  {\it anti-periodic} boundary conditions
for the gauge fields
 at infinity along $x^{-}$ and the vanishing ones along the
 $x^{1}$. These  boundary conditions fix also
the  residual
gauge invariance  with respect to the $x^{1}-$dependent
gauge transformations. For the scalar fields we assume vanishing
boundary conditions at infinity along the spatial coordinates.

The canonical Hamiltonian may then be written as

$$ H_{c}=\int d^2x \bigl[\,({\cal D}_{1}\phi)({\widetilde {\cal
D}}_{1}\phi^{*})-
A_{+} \Omega \, \bigr] \eqno(4) $$

\nl where

$$\Omega=ie(\pi\phi-\pi^{*}\phi^{*})+a \epsilon^{+ij}\partial_{i}A_{j}
+\partial_{i}\pi^{i}. \eqno(5)$$

\nl The primary constraints  following from (2) are

$$\eqalignno{ \pi^{+}&\approx 0 & (6a)\cr
{\top^{i}\equiv\pi^{i}-a\epsilon^{+ij} A_{j}} & \approx 0; \qquad \qquad i=-,1
&(6b)\cr
{\top\equiv \pi- {\widetilde {\cal D}}_{-}\phi^{*}} & \approx 0 &(6c)\cr
{\top^{*}\equiv \pi^{*}- {\cal D}_{-}\phi} &\approx 0 & (6d)\cr} $$

\nl and the preliminary Hamiltonian is

$$H'= H_{c}+ \int d^2x \bigl [ \, u\top+u^{*}\top^{*}+u_{i}\top^{i}+
u_{+}\pi^{+}\, \bigr ]
\eqno(7) $$

\nl where $u,u^{*},u^{i},u_{+}\,$ are lagrange multiplier fields.
We postulate initially
the standard canonical {\it equal-$\tau$}
Poisson brackets with the nonvanishing ones given by
$\{\pi^{\mu}(x),A_{\nu}(y)\}= -{\delta^{\mu}}_{\nu} \delta^{2}(x-y),\quad
\{\pi(x),\phi(y)\}=\{\pi^{*}(x),\phi^{*}(y)\}=-\delta^{2}(x-y)$, with
$x$ standing  for $(x^{-},x^{1})$ and $\tau$  suppressed for
convenience.
The following Poisson brackets among the
constraints are easily derived

$$\eqalignno{ \{\top^{i},\top^{j}\} &= -2a \epsilon^{+ij} \delta^2(x-y) &
(8a)\cr
\{\top^{i},\top \} &= -i {\delta^{i}}_{j} \phi^{*} \delta^2(x-y) &
(8b)\cr
\{\top^{i},\top^{*} \} &= i {\delta^{i}}_{j} \phi \delta^2(x-y) &
(8c)\cr
\{\top,\top^{*} \} &= -2 {\widetilde {\cal D}}^{x}_{-} \delta^2(x-y) &
(8d)\cr
\{\top,\top \} = \{\top^{*},\top^{*}\} &=0 &(8e) \cr}  $$

\nl while $\pi^{+}$ gives vanishing brackets  with all of them.
We make the {\it convention} that the first variable in an equal-$\tau$
bracket refers to the variable $x$
 while the second one to $y$. The  $\Omega$ generates
gauge transformations of the canonical variables
and  gives {\it weakly}
vanishing brackets with the
constraints (6) and $H'$.  The
evolution in $\tau$ of a dynamical variable  is
determined from $df(x,\tau)/d\tau\,= \{f(x,\tau),H'(\tau)\}+\partial f/
\partial \tau$.
Requiring the {\it persistency in $\tau $} of the
constraint $\,\pi^{+} \approx 0\,$ and noting that
$\{\pi^{+}(x,\tau),H'(\tau)\}\approx \Omega(x,\tau)$ we are led to a new
{\it secondary constraint }
$\Omega\approx 0$. The persistency  requirement for the others  results in
the consistency equations for  determining the lagrange multiplier
variables. We next go to the
extended Hamiltonian
$H''=H'+\int d^2x \,v\,\Omega\,$, where $v$ is a lagrange multiplier
and  repeat the procedure. We find that no new
secondary constraints are generated.
The  constraints $\pi^{+}\approx 0$ and
$\Omega \approx 0$ are first class
while the remaining ones
are  second class. The
{\it light-cone gauge} on the phase space
is defined by adding to the above set two {\it gauge-fixing constraints} $\,
A_{+}\approx 0$ and $A_{-}\approx 0$, since we have
two first class constraints. All the
constraints in the set now become second class.
The persistency of the gauge-fixing constraints is easily shown to be
secured implying that the gauge is  {\it accessible} and thus
there is no inconsistency in adopting the above two {\it local}
(weak) gauge-fixing conditions.
Similar arguments clearly hold also in the conventional Coulomb
gauge formulation.

We now  construct the Dirac bracket to
implement the above set of constraints in the theory.
The following   star bracket

$$\{f,g\}^{*}= \{f,g\}-\int d^2u \{f,\pi^{+}(u)\}\{A_{+}(u),g\}
            +\int d^2u \{f, A_{+}(u) \}\{ \pi^{+}(u),g\}$$

\nl has the property that it vanishes for arbitrary $g$ (or $f$)
when $f$ (or $g$) is equal to one of the variables $\pi^{+}$ or $A_{+}$.
We may then set $\pi^{+}=0, A_{+}=0$ as strong [18] equalities
and they are
{\it eliminated} from the theory. The star brackets of
the constraints are found to
coincide with the corresponding Poisson brackets and
the same holds true for the star brackets among the
remaining canonical variables. We may thus effectively ignore $A_{+}$
and $\pi^{+}$ completely and continue using Poisson brackets.
The modifications needed to take care
of the remaining set of constraints,  which we rename
 as $\top_{m},\quad m=1,2..6$:
$\top_{1}\equiv\top^{-},\top_{2}\equiv\top^{1},\top_{3}\equiv \top,
\top_{4}\equiv \top^{*}, \top_{5}\equiv A_{-}, \top_{6}\equiv \Omega\,$ is
done by following the standard procedure.

Define the constraint matrix $\,C (x,y)\,$

$$\Vert \{\top_{m},\top_{n}\}\Vert=
\left (\matrix {
0      & -2a    & -i\phi^{*}   & i \phi   & -1    & 0  \cr
2a     & 0      &  0           &  0       &  0    & 0  \cr
i \phi^{*}  &  0  & 0      & -2{\partial^{x}}_{-} & 0  & 0  \cr
-i \phi   &  0  & -2{\partial^{x}}_{-} & 0 & 0 &  0 \cr
1  & 0 & 0  &  0 & 0   &  - {\partial^{x}}_{-}  \cr
0 & 0 & 0 & 0 &  -{\partial^{x}}_{-} & 0 \cr } \right) \delta^2(x-y)
\eqno(9) $$

\nl The inverse
matrix is defined by

$$\int d^2z \,C_{mk}(x,z)\,{C^{-1}}_{kn}(z,y) = \delta_{mn}\delta^2(x-y)
\eqno(10)$$

\nl We find $\, C^{-1}(x,y)\,$ to be given by

$$ {\left (\matrix {
0      & -4 a{{\partial}^{x}}_{-} & 0   & 0   & 0   & 0  \cr
4 a {{\partial}^{x}}_{-}     &
[\phi^{*}(x)\phi(y)+\phi(x)\phi^{*}(y)]
   & {2ai}\phi(x)
    & - {2ai}\phi(x)^{*}      &  0    &  -{4 a}  \cr
0   &   {2ai}\phi(y)  & 0      &  (2a)^2 & 0  & 0  \cr
0  & - {2ai}\phi^{*}(y)  & (2a)^2  & 0 & 0 &  0 \cr
0  & 0 & 0  &  0 & 0   &  2(2a)^2  \cr
0 & -{4 a} & 0 & 0 & 2(2a)^2  & 0 \cr } \right)} {K(x-y)\over ({2a})^2}
\eqno(11) $$
\bigskip
\nl where $\,K(x-y)=-(1/4)\,\epsilon(x^{-}-y^{-})\,\delta(x^{1}-y^{1}), \,
{\partial_{-}}^{x} K(x,y)=(-1/2) \delta^2(x-y)\,$. Here
$\epsilon (x) = 1$ for $x>0$, $-1$ for
$x<0$ and $ \epsilon(0)=0$. The Dirac bracket which implements all the
constraints $\top_{m}$ is then constructed  to be

$$\{f,g\}_{D}= \{f,g\}-\int d^2u d^2v \, \{f,\top_{m}(u)\}\,
C^{-1}_{mn}(u,v)\,\{\top_{n}(v),g \} \eqno(12) $$
\bigskip
\nl It has the property $\{f,T_{m}\}_{D}=\{T_{m},f\}_{D}=0$
for arbitrary dynamical variable $f$.
We may then set $\top_{m}=0 \,$ as strong equalities
and the hamilton's eq. involves now the Dirac bracket in place of the
Poisson one.
The only {\it independent variables} left are $\phi$ and $\phi^{*}$ since
$T_{m}=0,\, A_{-}=0 $ lead to
$\pi=\partial_{-}\phi^{*}$, $\pi^{*}=\partial_{-}\phi$, $A_{-}=\pi^{1}=0$,
$\pi^{-}=a A_{1}$ while  (5) reduces to the Lagrange eq. (3c) which
determines $A_{1}$ in terms of the charge density $j^{+}$.

{}From (12) we compute

$$ \{\phi,\phi\}_{D}=0, \,\, \{\phi^{*},\phi^{*}\}_{D}=0,
\quad \{\phi,\phi^{*}\}_{D}= \{\phi^{*},\phi\}_{D}= K(x,y)\quad \eqno(13)$$

\nl which are the well known {\it light-front Dirac brackets}.
 Some other useful ones are

$$\eqalignno{ \{\pi,A_{1}\}_{D}&={-i\over {4a}}{\bigl [ -4 \pi(x) K(x,y)
+\phi^{*}\delta^2(x-y) \bigr ]} & (14a) \cr
\{\pi^{*},A_{1}\}_{D}&={i\over {4a}}{\bigl [ -4 \pi^{*}(x) K(x,y)
+\phi\delta^2(x-y) \bigr ]} & (14b) \cr
\{\phi,A_{1}\}_{D}&={i\over {2a}}{\bigl [ \phi(y) -
2 \phi(x)\bigr ] K(x,y)} & (14c) \cr
\{A_{1},A_{1}\}_{D}& = {1\over ({2a})^{2}}{\bigl [ \phi(x)\phi^{*}(y)+
\phi^{*}(x)\phi(y) \bigr] K(x,y)} & (14d) \cr }$$

\nl The light-front Hamiltonian in the light-cone gauge,
 obtained by substituting all the  constraints in $H''$,
  takes the simple form

$$H(\tau)=\int d^{2} x \, \,({\cal D}_{1}\phi)({\widetilde {\cal
D}}_{1}\phi^{*})
\eqno(15)$$

\nl to be compared with the involved one obtained in
the equal-time formulation [6-9].
There is still a $U(1)$ {\it global}
gauge symmetry generated by $Q$.
The scalar fields transform under this symmetry but they are left
invariant under the local gauge transformations since,
${ \{\Omega,f\}}_{D}=0 $.

It is mandatory to check the {\it self-consistency}. From the
hamilton's eq. for $\phi$ we derive  (we set $e=1$)

$$ \eqalignno {\partial_{-}\partial_{+}\phi(x,\tau)
  & =\{\pi^{*}(x,\tau),H(\tau)\}_{D}  \cr
 & = {1\over 2}{\cal D}_{1}{\cal D}_{1}\phi +
\int d^2y\, j^{1}(y,\tau)\,\{\pi^{*},A_{1}\}_{D} \cr
 &= {1\over 2}{\cal D}_{1}{\cal D}_{1}\phi +
 {1\over {4a}}\,\bigl [\,\phi\,(\phi^{*}{{\cal D}}_{1}\phi-
\phi{\widetilde{\cal D}}_{1}\phi^{*}) \,
 +\,\pi^{*}(x,\tau)\, \cr
 & \int d^2 y \,(\phi^{*}{{\cal D}}_{1}\phi-
\phi{\widetilde{\cal D}}_{1}\phi^{*})(y,\tau) \epsilon(x^{-}-y^{-})
\delta (x^{1}-y^{1})\, \bigr ] &(16)\cr }$$

\nl Comparing  (16) with (3a) it is suggested to
introduce
a new variable in our formulation, indicated for convenience by
(the above eliminated) $A_{+}$, and whose expression is that obtained from
solving (3d). Eq. (3b) is derived from  $\Omega=0$ and it is then
 straightforward to check
(3c). The hamiltonian theory in the light-cone gauge constructed here
is thus  shown  self-consistent.
The variable $A_{+}$ has
{\it reappeared} and we are {\it effectively} imposing
 $A_{-}=0 $ and {\it not}
$A_{\pm}=0$ which would in its turn imply setting
the gauge invariant quantity $F_{+-}$
to be vanishing, leading in general to  contradiction with (3d).
Similar discussion can be made in the Coulomb gauge formulation as regards
to $A^{0}$.  Contrary
to the suggestions made in [9] there arises
{\it no inconsistency  on using the
non-covariant local gauges}.
That
only the nonlocal gauges may describe [9] the fractional statistics
consistently in the present theory is not tenable.
The conventional  Coulomb gauge or nonlocal
gauge-fixing  conditions lead to quite
complicated interactions and hamiltonian  and renormalized
theory seems difficult to construct.
They do have the advantage  of showing
a dual description [6-9] in terms of free fields with
multivalued operators and the  manifest  fractional statistics which
arises from the graded equal-time commutation relations.
In our case also it is possible to rewrite the
Hamiltonian density in (15) as $ {\cal H}=(\partial_{1}\hat\phi)
(\partial_{1}\hat \phi^{*})$ if we note that $A_{1}=\partial_{1}\Lambda$
where
$\,8a\,\Lambda(x^{-},x^{1})=\int d^2y \,\epsilon(x^{-}-y^{-})\,
\epsilon(x^{1}-y^{1})\,
j^{+}(y)\,$ and define $\,{\hat\phi}=e^{i\Lambda}\phi\,$,
 $\,{\hat\phi^{*}}=e^{-i\Lambda}\phi^{*}\,$.
 In view of (14c-d) the
field $\hat\phi$ then does not have the vanishing Dirac
bracket (or commutator) with itself.
The theory is quantized via the correspondence  of
$i\{f,g\}_{D}$ with the commutator $[f,g]$ among the corresponding
field theory operators. When there is ambiguity in the
operator ordering we resort
to the Weyl ordering.

\bigskip
\medskip
\nl {\bf 3. Relativistic Covariance and Absence of Anomaly}
\bigskip

In the non-covariant gauge
like the one chosen here  the manifest covariance is lost and even the scalar
field may acquire some
unconventional transformation properties.  The
relativistic invariance is shown by constructing the field theory
space time symmetry generators and verifying that they give rise to
Poincar\'e algebra.
The canonical energy-momentum tensor derived from (1) is given by

$${\theta_{c}}^{\mu\nu} = ({\widetilde {\cal D}}^{\mu}\phi^{*})
{(\partial^{\nu}\phi)}+
({\cal D}^{\mu}\phi)(\partial^{\nu}\phi^{*})
+a \epsilon^{\sigma\mu \rho}A_{\sigma}\partial^{\nu}A_{\rho}-\eta^{\mu\nu}
{\cal L} \eqno(17)$$

\nl where $\partial_{\mu}{\theta_{c}}^{\mu \nu}=0\,$ by construction.
In the light-cone gauge  they get simplified, for example,

$$\eqalignno {{\theta_{c}}^{++} &=  2\,\pi\pi^{*} &(18a) \cr
{\theta_{c}}^{+1} &= -(\pi\partial_{1}\phi+\pi^{*}\partial_{1}\phi^{*} )
&(18b) \cr
{\theta_{c}}^{+-} &=  \,({\cal D}_{1}\phi)({\widetilde {\cal D}}_{1}\phi^{*})
={\cal H} &(18c) \cr }$$

\nl The momentum generators defined by $P^{\mu}= \int d^2x \,
{\theta_{c}}^{+\mu}\,$  are conserved and shown  to
generate the  translations, e.g., $\{\phi,P_{\mu}\}_{D}=
\partial_{\mu}\phi,\,\,\{\phi^{*},P_{\mu}\}_{D}=
\partial_{\mu}\phi^{*}\,$ when we make use of the
boundary conditions.
The invariance of the classical lagrangian (1) under Lorentz
transformation results [19] in the following conserved current

$$\eqalignno {J^{\mu\rho\sigma} &= -J^{\mu\sigma\rho} \cr
&= x^{\rho}{\theta_{c}}^{\mu\sigma}-x^{\sigma }{\theta_{c}}^{\mu\rho}
-i {\partial {\cal L}\over {\partial (\partial_{\mu}A_{\alpha}})}
{{(\Sigma^{\rho\sigma})}^{\alpha}}_{\beta} A^{\beta} & (19) \cr }$$

\nl where $\partial_{\mu}J^{\mu\rho\sigma}=0\,$ on the mass shell and
$\,{(\Sigma_{\rho\sigma})}_{\alpha\beta}=i(\eta_{\rho\alpha}\eta_{\sigma\beta}
-\eta_{\rho\beta }\eta_{\sigma\alpha})$ .
The generators of the Lorentz transformation
on the light-front may hence be defined by

$$\eqalignno {M^{\mu\nu}=-M^{\nu\mu}&= \int d^2x \, J^{+\mu\nu} \cr
&= \int d^2x \, \bigl [ x^{\mu} {\theta_{c}}^{+\nu}
-x^{\nu} {\theta_{c}}^{+\mu}\,- ( A^{\mu}\pi^{\nu}- A^{\nu} \pi^{\mu})
  \bigr ] &(20) \cr}$$

\nl and in the light-cone gauge they simplify to

$$\eqalignno { M^{-1} &= \int d^2x \, \bigl [ x^{-} {\theta_{c}}^{+1}
-x^{1} {\theta_{c}}^{+-}\,- a A_{1}^{2} \bigr ] &(21a)\cr
M^{+1} &= x^{+}P^{1}- \int d^2x \,
x^{1} {\theta_{c}}^{++}  &(21b) \cr
M^{+-} &= x^{+} P^{-} - \int d^2x \,
x^{-} {\theta_{c}}^{++}  & (21c) \cr} $$

\nl The  expressions of the generators
as obtained on
using the symmetric Belinfante tensor  [20,19],
$\,{\theta_{B}}^{\mu\nu}=[\, {\theta_{c}}^{\mu
\nu}+ a \epsilon^{\lambda\mu\beta}
\partial_{\lambda}(A_{\beta}A^{\nu})]\,$,   or the
symmetric gauge invariant one [6]
differ from (22) only by a surface term.
It is to be stressed that the generators (21) and those for
$P^{\mu}$ above are  {\it perfectly legitimate set } [19]
to use in order to check  the relativistic invariance
of the theory  under discussion.
We showed already that
the lagrange eqs. (3) are recovered in the hamiltonian framework. A direct
verification of the closure of the Poincar\'e algebra on the mass shell,
e.g., where we use (3), then
becomes straightforward, though tedious. We note, for example,

$$\eqalignno {\{{\theta_{c}}^{++},{\theta_{c}}^{+1}\}_{D}&=
-\pi(x)[(\partial_{1}\phi(y)){\partial^{x}}_{-}\delta^{2}(x-y)
+\pi^{*}(y){\partial^{x}}_{1}\delta^2(x-y)] \cr
&\quad -\pi^{*}(x)[(\partial_{1}\phi^{*}(y)){\partial^{x}}_{-}\delta^{2}(x-y)
+\pi(y){\partial^{x}}_{1}\delta^2(x-y)], &(22a) \cr}$$

$$\eqalignno{\{{\theta_{c}}^{++},{\theta_{c}}^{+-}\}_{D}& =
\;2\pi(x)\{\pi^{*},{\cal D}_{1}\phi{\widetilde{\cal D}}_{1}\phi^{*}\}_{D}+
2\pi^{*}(x)\{\pi,{\cal D}_{1}\phi{\widetilde{\cal D}}_{1}\phi^{*}\}_{D} \cr
&= 2a\, (\partial_{-}A_{1})(\partial_{-}A_{+})\delta^2(x-y)-\pi(x)
({\cal D}_{1}\phi(y)){{\widetilde{\cal D}}^{y}}_{1}\delta^2(x-y) \cr
&\quad - \pi^{*}(x)({\widetilde{\cal D}}_{1}\phi^{*}(y))
{{\cal D}^{y}}_{1}\delta^2(x-y). &(22b)\cr } $$

$$\eqalignno {\{{\theta_{c}}^{++}, {A^{2}}_{1}\}_{D} &=
-2 A_{1}(\partial_{-}A_{1}){\delta^{2}(x-y)}, &(22c) \cr
 ({\theta_{c}}^{1-}-{\theta_{c}}^{-1})  &=
\,a\,[\partial_{1}(A_{+}A_{1})-
\partial_{+}{A^{2}}_{1}-\partial_{-}{A^{2}}_{+}], &(22d) \cr} $$

\nl The light-front hamiltonian formulation of (1)
in the light-cone gauge is found  to be  Poincar\'e invariant.

The  independent variables $\phi,
\phi^{*}\,$  satisfy the light-front Dirac brackets (13) and
the expressions
(19) and (21a) differ from those of the free field
theory due to the contributions of the now dependent  field $A_{1}$.
Such  extra terms have been called [10,6]  {\it anomalous spin}
induced on the scalar field
due to the constrained dynamics generated  by the Chern-Simons term.
They may also not be removed
by a redefinition of the generators which are required to satisfy
the Poncar\'e algebra for relativistic invariance. We discuss this now
in our context carefully by
considering the Lorentz transformation of the scalar field.
{}From (3), (13), (14), and (21)  we derive after some algebra

$$ \eqalignno {\{\phi(x,\tau), M^{-1}(\tau)\}_{D}& =
[x^{-}\partial^{1}-x^{1}\partial^{-}]
\phi(x,\tau) \cr
&- {i\over 2} \phi(x,\tau) \int d^2y\, \epsilon(x^{-}-y^{-})
\delta (x^{1}-y^{1})\, A_{1}(y,\tau)\, &(23a)} $$

$$ \{\phi(x,\tau), M^{+1}(\tau)\}_{D}= [x^{+}\partial^{1}-x^{1}\partial^{+}]
\, \phi(x,\tau), \eqno(23b) $$

$$ \{\phi(x,\tau), M^{+-}(\tau)\}_{D}= [x^{+}\partial^{-}-x^{-}\partial^{+}]
\, \phi(x,\tau) \eqno(23c),  $$

\nl and on combining (23a) and (23b)  we get

$$\eqalignno { \{\phi(x,\tau), M^{21}(\tau)\}_{D}&=
[x^{2}\partial^{1}-x^{1}\partial^{2}] \,
\phi(x,\tau)\cr
& + {i\over 2} \phi(x,\tau) \int d^2y\, \epsilon(x^{-}-y^{-})
\delta (x^{1}-y^{1})\, A_{1}(y,\tau)\, &(23d) \cr} $$

The extra second term in (23a) or (23d) has been called [10,6] in the
the conventional treatment
an {\it anomaly} arising from  the anomalous spin
term in the generator $M^{-1}$ or $M^{12}$. Its presence, however, does not
lead to the breakdown of the closure of the space time generators to the
Poincar\'e algebra.
{}From our discussion, however, it is clear that we may as well interpret
the anomalous transformation (23a) or (23d) here or those
found in the previous works [10,6]  as {\it gauge artefacts}.
It is clear  from the discussion here that, for example,
the unusual transformations of
$A_{-}$ under space time rotations, viz, $\{M^{\mu\nu},A_{-}\}_{D}=0$ or
under translations, $\{P^{\mu},A_{-}\}_{D}=0$,
originate from the  construction of the Dirac bracket (12).
If $A_{-}$ transformed
normally we would be led in the light-cone gauge to $A_{1}=0$ as well.
The unusual (anomalous)
transformation above may  not be thus considered as
totally unexpected in the non-covariant gauge being used.
This  is reinforced by the verification of that
the components of the {\it gauge invariant vector} $j^{\mu}$, whether defined
in terms of the scalar fields or in terms of the gauge fields according to
(3), continue to possess the usual transformation
properties of vector field
and no anomalous terms are generated.
 To illustrate, we find

$$\{A_{1}(x,\tau),M^{-1}(\tau)\}_{D}=(x^{-}\partial^{1}-x^{1}\partial^{-})A_{1}
-A_{+}+{1\over {\partial_{-}}}\,{\partial_{1}A_{1}}.\eqno(24)$$

\nl The last term on the right hand side is to be considered as
gauge artefact rather than an anomalous term and it is correctly absent from

$$\{\partial_{-}A_{1}(x,\tau),M^{-1}(\tau)\}_{D}=
(x^{-}\partial^{1}-x^{1}\partial^{-})(\partial_{-}A_{1})
-(\partial_{-}A_{+})\eqno(25)$$

\nl since $j^{+}\sim \partial_{-}A_{1}$ and $j^{1}\sim \partial_{-}A_{+}$
and $j^{\mu}$ is a contravariant vector.
There is no
genuine anomaly in the behavior of the scalar field (or the field $A_{1}$)
under the Lorentz
transformations which agrees  with the same result obtained in the
recent gauge independent discussions [12,13] in the equal-time formulation.
The physical outcome, e.g., the
emergence of (anyonic) excitations obeying fractional
statistics should emerge
in the dynamics of the relativistic theory here as described
by (13) and (15), and not be considered as a consequence of the
unconventional
transformation law (gauge artefact) of the scalar field
connected with the gauge-fixing
conditions, which may  as well be nonlocal and non-linear. A
similar  discussion  may  be given  in the conventional local
Coulomb gauge.
Because of its simplicity the light-front quantized
theory in the light-cone gauge
promises to be  a useful framework for discussing  renormalization of the
model (1).

\bigskip

\bigskip

\centerline {\bf Acknowledgements}

\bigskip
The author would like to thank Professor Abdus Salam,
the International Atomic Energy
Agency and UNESCO for the hospitality at the International
Centre for Theoretical Physics, Trieste.
Acknowledgements are due to  F. Caruso, R. Shellard, and
B. Pimentel for comments.

\bigskip

\nl {\bf References:}

\bigskip
\item {[1]}S. Deser, R. Jackiw, and S. Templeton, Phys. Rev. Lett. {\bf 48}
(1982) 975; Ann. Phys. (N.Y.) {\bf 140} (1982) 372.
\item{[2]}R. Jackiw, Topics in planar physics, MIT preprint No. MIT-CTP-
1824 (1989), unpublished; R. Mackenzie and F. Wilczek, Int. J. Mod.
Phys. {\bf A 3} (1988) 2827.
 \item {[3]}R.B. Laughlin, Phys. Rev. Lett.
{\bf 50} (1983) 1395; {\bf 60} (1988) 2677.
\item {[4]}F. Wilczek, ed.,  {\it Fractional Statistics and Anyon
Superconductivity} (World Scientific, Singapore, 1990).

\item {[5]} see for example,
 S. Forte, Rev. Mod. Phys. {\bf 64} (1992) 193 and references therein.

\item {[6]} G.W. Semenoff, Phys. Rev. Lett. {\bf 61} (1988) 517; G.W.
Semenoff and P. Sodano, Nucl. Phys. {\bf B 328} (1989) 753.
\item {[7]} T. Matsuyama, Phys. Letts. {\bf B 228} (1989) 99.
\item{[8]} A. Foerster and H.O. Girotti, Phys. Lett. {\bf B 230} (1989)
83.

\item {[9]} R. Banerjee, A. Chatterjee, and V.V. Sreedhar, Ann. Phys.
(N.Y.) {\bf 222} (1993) 254.

\item {[10]} C. Hagen, Ann. Phys. (N.Y.) {\bf 157}(1984) 342. See also
H. Shin, W.T. Kim, J.K. Kim, and Y.J. Park, Phys. Rev. {\bf D46}
 (1992) 2730.
\item {[11]} See for example,
P.P. Srivastava, Nuovo Cimento {\bf 64 A} (1981) 259.
\item {[12]} R. Banerjee, Phys. Rev. {\bf D48} (1993) 2905.
\item{[13]} R. Amorim and J. Barcelos-Neto, UFRJ, Rio de Janeiro preprint,
IF-UFRJ-05/94.
\item {[14]}P.A.M. Dirac, Rev. Mod. Phys. {\bf 21} (1949) 392.

\item{[15]}K.G. Wilson, Nucl. Phys. B (proc. Suppl.) {\bf 17} (1990);
R.J. Perry, A. Harindranath, and K.G. Wilson,
Phys. Rev. Lett. {\bf 65} (1990)  2959.
\item{[16]}S.J. Brodsky and H.C. Pauli, {\it Schladming Lectures}, {\sl SLAC}
 preprint {\sl SLAC}-PUB-5558/91.
\item{[17]}P.P. Srivastava,
{\it Higgs mechanism ({\sl tree level}) in light-front
quantized field theory},  Ohio State University preprint 92-0012,
 SLAC HEP data base no.  PPF-9202, December 91; Nuovo Cimento {\bf A 107}
 (1994) 549;
\item{} Lectures on {\it  Light-front quantized field theory:}
{\sl Spontaneous symmetry breaking. Phase transition in $\phi^{4}$ theory},
Proceedings {\it XIV Encontro Nacional de Part\|culas e Campos, Caxambu, MG},
1993, pgs. 154-192, {\it
Sociedade Brasileira de F\|sica}. Available
also from {\it hep-th@xxx.lanl.gov}, no.  9312064.

\item{[18]}P.A.M. Dirac, {\it Lectures
in Quantum Mechanics}, Benjamin, New York, 1964; E.C.G. Sudarshan and
N. Mukunda, {\it Classical Dynamics: a modern perspective}, Wiley, N.Y.,
1974.

\item {[19]} P.P. Srivastava, Nucl. Phys. {\bf B
64} (1973) 499.
\item {[20]} F.J. Belinfante, Physica {\bf 7} (1940) 449.

\bye